\documentclass[twocolumn,showpacs,showkeys,amssymb,prd]{revtex4}

\usepackage{graphicx}% Include figure files
\usepackage{dcolumn}% Align table columns on decimal point
\usepackage{bm}% bold math
\usepackage{epsfig}
\begin{document}
\newcommand{\gsim}{\hbox{\rlap{$^>$}$_\sim$}}
\newcommand{\lsim}{\hbox{\rlap{$^<$}$_\sim$}}

\title{Neutrinos And Cosmic Rays From Gamma Ray Bursts}

\author{Arnon Dar}
\affiliation{Department of Physics, Technion, Haifa 32000, Israel}

\begin{abstract}
The upper limit on the flux of ultra high energy neutrinos from 
$\gamma$-ray bursts (GRBs)  reported recently by the IceCube 
collaboration contradicts 
predictions based on the Fireball model of GRBs, but does not exclude 
GRBs as a main source of ultra-high energy cosmic rays.
\end{abstract}

\maketitle

In a recent letter published in Nature$^1$, the IceCube collaboration 
reported an experimental upper limit on the flux of ultra-high energy 
(energies above EeV$=10^{18}$ electronvolts) neutrinos from $\gamma$-ray 
bursts (GRBs) that is at least a factor of 3.7 below theoretical 
predictions$^{2,3,4}$ based on the fireball model of GRBs. Hence, they 
concluded that GRBs are not the only source of ultra-high energy (UHE) 
cosmic rays or the efficiency of UHE neutrino production is much lower 
than that predicted. However, while presenting important experimental 
results, the IceCube collaboration has over-interpreted their results.

The fireball model of GRBs that was used to predict the expected 
fluxes of ultra-high energy cosmic rays and neutrinos$^{2,3,4}$ from GRBs, 
has already been challenged by many observations of GRBs and their 
afterglows$^{5}$. Neutrino flux estimates based on this model 
cannot be relied on in drawing any conclusions either on cosmic ray 
acceleration to ultra-high energy in GRBs or on the production 
of UHE neutrinos in GRBs.  In particular,
alternative estimates of the fluxes of UHE cosmic rays and 
neutrinos from GRBs$^6$ that were based on the cannonball model of GRBs, 
which, so far, was shown to reproduce well the  observed properties 
of GRBs and their afterglows$^{7,8}$, have yielded neutrino fluxes 
that are much smaller than the upper limit obtained by the IceCube 
collaboration$^{1}$.

Moreover, recent studies with the Pierre Auger Observatory (PAO) of the 
nuclear mass composition of ultra-high energy cosmic rays that reach Earth 
indicate that their composition changes gradually between 4 EeV and 40 EeV 
from proton-dominated composition to iron-dominated composition$^9$ 
(provided that standard high energy particle physics is still valid at 
ultra-high energies). The production of neutrinos with ultra-high energy 
$E$ requires accelelaration in GRB fireballs of protons and/or other 
atomic nuclei of mass number $A$ to energies $\gsim 10\, A\,E$. But 
complex nuclei cannot be accelerated in the alleged GRB fireballs to such 
ultra-high energies because they disintegrate in collisions with fireball 
photons long before they reach these ultra-high energies.

The flux of UHE extragalactic cosmic ray nuclei is strongly reduced by 
photo-disintegration in collisions with photons of the cosmic infra-red 
and microwave background radiation$^{6,10}$. 
Hence, if the PAO composition is correct$^{11}$ then
extragalactic sources, 
including extragalactic GRBs, cannot be the main source of the UHE cosmic 
ray nuclei observed in the Galaxy, and$^{12}$:
(a) the 'ankle' near 4 EeV in the energy spectrum of Galactic cosmic rays 
most probably is the energy beyond which the deflections of CR protons and 
He4 nuclei in the Galactic magnetic fields can no longer isotropise them 
nor prolong significantly their escape from the Galaxy, and (b) the 
spectral break near 50 EeV observed by the Fly's Eye High Resolution 
(HiRes) experiment$^{13}$ and by PAO$^{14}$ may be the 'escape-break' of 
ultra-high energy iron nuclei from the Galaxy, whose energy is $Z/2=13$ 
times higher than that of the helium escape-break rather than the so 
called 'GZK 
cutoff' - the 
effective threshold for energy losses of cosmic ray protons by pion 
production in collisions with the cosmic microwave background (CMB) 
radiation. These losses exponentially suppress  the 
extragalactic fluxes of 
cosmic rays beyond $\sim 50\, A$ EeV, as noted by Greisen$^{15}$ and by 
Zatsepin and Kuzmin$^{16}$ in 1966 right after the discovery of the CMB.

Galactic GRBs, most of which,  mercifully$^{17}$, are 
beamed away from Earth, can be the main source of Galactic cosmic rays at 
all energies$^{12,18}$.


\begin{thebibliography}{99}

\bibitem[Abbasi et al., 2012]{Abbasi2012}%
IceCube Collaboration: Abbasi, R. et al.
An Absence of Neutrinos Associated with Cosmic Ray Acceleration in 
Gamma-Ray Bursts. {\it Nature} {\bf 484}, 351-354 (2012).
[arXiv:1204.4219]

\bibitem[Waxman, 1995]{Waxman1995}%
Waxman, E. Cosmological gamma-ray bursts and the highest
energy cosmic rays. {\it Physical Review Letters} {\bf 75}, 386-389 
(1995).  [arXiv:astro-ph/9505082]

\bibitem[Waxman and Bahcall, 1997]{Waxman1997}%
Waxman, E. \& Bahcall, J. High energy neutrinos from cosmological
gamma-ray burst reballs. {\it Physical Review Letters} {\bf 78}, 2292-2295 
(1997). [arXiv:astro-ph/9701231]


\bibitem[Guetta et al., 2004]{Guetta2004}%
Guetta, D., Hooper, D., Alvarez-Mu~niz, J., Halzen, F. \&
Reuveni, E. Neutrinos from individual gamma-ray bursts in the
BATSE catalog. {\it Astroparticle Physics}  20, 429-455 (2004).
[arXiv:astro-ph/0302524]
 
\bibitem[Margutti et al., 2012]{Margutti2012}%
Margutti R., et al. 
The prompt-afterglow connection in Gamma-Ray Bursts: a comprehensive 
statistical analysis of Swift X-ray light-curves. 
eprint, arXiv:1203.1059 (2012) and references therein. See also, e.g.,
references [7],[8], and references therein. 

\bibitem[Dar and DeRujula, 2008]{Dar2008}%
 Dar, A. \&  De R\'ujula, A. Theory of Cosmic Rays. 
{\it Physics Reports}  {\bf 466}, 179-241 (2008). [arXiv:hep-ph/0606199]


\bibitem[Dado et al., 2009a]{Dado2009a}%
Dado, S.,  Dar A. \&  De R\'ujula, A.
The diverse broad-band light-curves of Swift GRBs reproduced with the 
cannonball model. 
{\it The Astrophysical Journal} {\bf 696}, 994-1020 (2009).  
[arXiv:0809.4776]

\bibitem[Dado et al., 2009b]{Dado2009b}%
Dado, S.,  Dar A. \&  De R\'ujula, A.
Short Hard Gamma Ray Bursts And Their Afterglows.
{\it  The Astrophysical Journal} {\bf  693}, 311-328 (2009). 
[arXiv:0807.1962]


\bibitem[Cazon et al., 2012]{Cazon2012}%
The Pierre Auger Collaboration, Cazon, L. et al. 
Studying the nuclear mass composition of Ultra-High Energy Cosmic Rays 
with the Pierre Auger Observatory, arXiv:1201.6265 
Proceedings of the 12th International Conference on Topics in 
Astroparticle and Underground Physics, TAUP 2011, Munich, Germany.
[arXiv:1201.6265]

\bibitem[Tkaczyk et al., 1975]{Tkaczyk1975}%
Tkaczyk, W.,  Wdowczyk, J. \& W.Wolfendale, A. 
Extragalactic heavy nuclei in cosmic rays. {\it J. Phys.
A.} {\bf  8},  1518-1529 (1975); Puget, J. L.,  Stecker, F. W. \& 
Bredekamp, J. H. Photonuclear interactions of ultrahigh energy cosmic rays 
and their astrophysical consequences.
 {\it  Astrophys. J.}  {\bf 205}, 638-654 (1976).


\bibitem[Abbasi et al., 2010]{Abbasi2010}%
The Fly's Eye HiRes Collaboration; R. U. Abbasi, et al.
Indications of Proton-Dominated Cosmic-Ray Composition Above 1.6 EeV.
{\it Physical Review Letters},{\bf 104}, 161101-161105 (2010).
[arXiv:0910.4184]

\bibitem[Dado et al., 2012]{Dado2012}%
Dado, S.  Dar, A. \& De R\'ujula, A. 
Origin of the ultra-high energy cosmic rays and their spectral break 
{\it Il Nuovo Cimento C} {\bf 35}, 01-08 (2012). 
 [arXiv:1011.2672]

\bibitem[Abbasi et al., 2008]{Abbasi2008}%
Fly's Eye HiRes  Collaboration;  Abbasi, R. U. et al.
First observation of the Greisen-Zatsepin-Kuzmin suppression. 
{\it Physical Review Letters} {\bf 100}, 101101-101105 (2008).

\bibitem[Abraham et al., 2010]{Abraham2010}%
Pierre Auger Collaboration; Abraham, J. et al.
Measurement of the energy spectrum of cosmic rays above 10$^{18}$ eV  
using the Pierre Auger Observatory.
{\it Physics Letters B} {\bf 685}, 239-246 (2010). 


\bibitem[Greisen, 1966]{Greisen1966}%
Greisen, K.  End to the Cosmic-Ray Spectrum? 
{\it Physical Review Letters} {\bf 16}, 748-750 (1966). 
 

\bibitem[Zatsepin and Kuzmin, 1966)]{Zatsepin1966}%
Zatsepin, G. T. \& Kuzmin, V. A. {\it 
Upper Limit of the Spectrum of Cosmic Rays.
Journal of Experimental and Theoretical Physics Letters} 
{\bf 4}, 78-80 (1966).

\bibitem[Dar et al., 2009]{Dar2009}% 
Dar, A., Laor, A. \&  Shaviv, N. J. Life Extinctions By Cosmic Ray Bursts.
{\it Physical Review Letters}, {\bf 80} 5813-5816 (1998)
[arXiv:astro-ph/9705008]. See also  Dar, A. \& De R\'ujula, A.
The threat to life from Eta Carinae and gamma ray bursts
{\it Astrophysics and Gamma Ray Physics in Space (eds.
A. Morselli and P. Picozza), Frascati Physics Series} {\bf Vol. XXIV},
513-523 (2002) [arXiv:astro-ph/0110162].


\bibitem[Dar and Plaga, 1999]{Dar1999}%
Dar, A. \& Plaga, R.
Galactic gamma-ray bursters - an alternative source of cosmic rays at all 
energies.  {\it Astronomy \& Astrophysics} {\bf 349}, 259-266 (1999).
 [arXiv:astro-ph/9902138]

\end{thebibliography}
\end{document}